\documentclass[12pt]{article}
\usepackage{graphicx}

\begin{document}

\title{Interaction-free measurement with mesoscopic devices on a GaAs/AlGaAs heterostructure}

\author{Hiroo Azuma\thanks{Email: hiroo.azuma@m3.dion.ne.jp}
\\
\\
{\small Advanced Algorithm \& Systems Co., Ltd.,}\\
{\small 7F Ebisu-IS Building, 1-13-6 Ebisu, Shibuya-ku, Tokyo 150-0013, Japan}\\
}

\date{\today}

\maketitle

\begin{abstract}
In this paper, we propose a method for implementing the interaction-free measurement (IFM)
with mesoscopic devices fabricated on a high mobility
\\
GaAs/AlGaAs heterostructure,
where a two-dimensional electron gas (2DEG) system is buried,
at low temperature.
The IFM is amazing evidence of nonlocality of the quantum mechanics
because the IFM offers us the capability to detect the existence of an object without interaction with it.
A scheme of Kwiat et al. for realizing the IFM lets a probability
that we fail to recognize the presence of the object be as small as we like,
using the quantum Zeno effect.
Constructing an interferometer of Kwiat et al. for the IFM,
we make use of techniques of electron quantum optics and
tunnelling microscopes
for the 2DEG.
We examine a shot noise of an electric current flowing through the interferometer.
Our method for performing the IFM aims at achieving a milestone
of quantum information processing in mesoscopic systems.
\end{abstract}

\bigskip


\bigskip

\section{\label{section-introduction}Introduction}
In this paper, we propose a method for implementing the interaction-free measurement (IFM)
with mesoscopic devices.
In 1981, Dicke proposed an early concept of the IFM~\cite{Dicke1981}.
Elitzur and Vaidman starts modern discussion about the IFM by taking up the following problem~\cite{Elitzur1993,Vaidman2001}:
``Suppose there is an object such that any interaction with it leads to an explosion.
Can we locate the object without exploding it?"

For example,
we assume the object that absorbs an interrogating photon with strong interaction
if the photon comes near the object in distance.
We want to examine whether or not the object exists in a black box without its absorption of the interrogating photon.
Elitzur and Vaidman show that the Mach-Zehnder interferometer works as implementation of the IFM.
However, their scheme is very simple and a probability of detecting the object without its absorption
for one interrogation is equal to $1/4$.
If repetition of interrogations is allowed,
the probability becomes $1/3$.
Moreover,
if the reflectivity of the beam splitter is adjusted,
it is improved and reaches almost $1/2$.

Kwiat et al. find a more refined experimental method for realizing the IFM~\cite{Kwiat1995}.
They put the absorbing object in an interferometer which consists of many beam splitters,
and inject a photon into it to examine whether or not the object exists.
In the scheme of Kwiat et al.,
the quantum Zeno effect plays an important role and its efficiency,
namely its detection rate for the absorbing object, approaches unity asymptotically as the number of the beam splitters grows.
In Reference~\cite{Kwiat1999},
this scheme is actually performed with high efficiencies of up to $73$\%
in experiments.

Moreover, Paul and Pavi{\v c}i{\'c} propose another experimental setup for the IFM.
Making use of the Fabry-P{\'e}rot interferometer,
they let their own scheme be very practical~\cite{Paul1997}.
Paul and Pavi{\v c}i{\'c}'s scheme is experimentally demonstrated by two research groups
operating in Sweden and Japan with efficiencies around $80$\% and $88$\%,
respectively~\cite{Tsegaye1998,Namekata2006}.

As mentioned above,
many researchers in the field of quantum information science are interested
in an experimental demonstration of the IFM.
This is because the IFM is not only a strange aspect of the quantum mechanics
but also an important element in quantum information processing.
If we obtain skill in executing the perfect IFM,
we can perform various operations of quantum information processing.
For example, Pavi{\v c}i{\'c} points out that techniques for the IFM improve
the atom-photon controlled-NOT gate~\cite{Pavicic2007}.
Azuma discusses how to take the Bell-basis measurement and
how to construct the controlled-NOT gate by using the IFM~\cite{Azuma2003,Azuma2004}.
In Reference~\cite{Azuma2006},
Azuma examines the IFM with an imperfect absorber.

So far,
in almost all setups of the IFM,
quantum interrogation of the absorbing object has been performed with a photon.
Most researchers prefer the photon for the quantum interrogation
because
it is suitable to construct the interferometer
and pieces of equipment needed for building the interferometer, such as beam splitters,
mirrors, and so on, are generally available.

However,
some attempts to utilize an electron as the interrogating particle for the IFM are undertaken.
In Reference~\cite{Strambini2010},
the Aharonov-Bohm ring with asymmetric electron injection is employed for detection of electron dephasing,
and this phenomenon is interpreted as the IFM.
In Reference~\cite{Chirolli2010},
the IFM based on the integer quantum Hall effect is proposed.

If we obtain the ability to accomplish the IFM with the interrogating electron,
we can apply the quantum information processing to quantum bits implemented with electrons in a fluent manner.
We can construct quantum information processors from semiconductor devices.
This is the primary motivation of the current paper.

In this paper, we think about a two-dimensional electron gas (2DEG) system,
which appears in a high mobility GaAs/AlGaAs heterojunction
at low temperature~\cite{Mimura1980,Ferry1997}.
If we perform experiments at near absolute zero temperature, $T\simeq 0$ K,
the mobility of the confined electrons being free to move in two dimensions is
on the order of $\mu=e\tau/m^{*}\simeq 1.0\times 10^{2}$ $\mbox{m}^{2}\mbox{V}^{-1}\mbox{s}^{-1}$.
Because the effective mass of the electron in GaAs is given by $m^{*}=0.067 m_{\mbox{\scriptsize e}}$,
its averaged relaxation time is estimated at $\tau\simeq 3.81\times 10^{-11}$ s around.

Moreover, the Fermi energy of GaAs is given by $E_{\mbox{\scriptsize F}}=0.014$ eV
and we obtain the Fermi velocity
$v_{\mbox{\scriptsize F}}=\sqrt{2E_{\mbox{\scriptsize F}}/m^{*}}\simeq 2.71\times 10^{5}$ ms${}^{-1}$,
so that the elastic mean free path of the electron in GaAs is approximately equal to
$l=v_{\mbox{\scriptsize F}}\tau\simeq 1.03\times 10^{-5}$ m.
Thus, if we fabricate a nanostructure system
whose typical length is smaller than the elastic mean free path $l$ on the GaAs/AlGaAs heterojunction,
particles move in the active region without scattering.
This ballistic transport is one of the astonishing characteristics of the mesoscopic system.

To perform the scheme of Kwiat et al. for the IFM in the 2DEG system,
we make use of techniques of electron quantum optics for implementation of the single-electron source
and the beam splitter~\cite{Feve2007,Mahe2010,Bocquillon2012,Roussel2017}
and the tunnelling microscope for realizing the absorbing object~\cite{Binnig1982}.
The electron quantum optics is a particular perspective on electronic ballistic transport in quantum conductors
and it aims a counterpart  of orthodox photon quantum optics.
These topics belong to the mesoscopic physics,
which treats quantum transport of electrons and holes
confined in one or two dimensions.

In the interferometer of Kwiat et al.,
we let electrons travel along a quantum wire one by one with a constant time interval.
Moreover,
we replace the absorbing object with a probe of the tunnelling microscope.
Thus,
varying a voltage applied to the tip of the microscope,
we can adjust a probability that the probe captures the electron with ease.
We examine a shot noise of the electric current flowing
through the interferometer~\cite{Imry1997,Buttiker1990,Martin1992,Buttiker1992,deJong1997,Blanter2000}.

Detection of the 2DEG with the tunnelling microscope is experimentally realized in Reference~\cite{Topinka2001}.
Direct observation of the Landau quantization with the tunnelling spectroscopy is reported in References~\cite{Cheng2010,Yan2012}.
Hence, we can expect the tunnelling microscope to detect the 2DEG through a thin layer of AlGaAs on the heterojunction.

This paper is organized as follows.
In Section~\ref{section-review-IFM}, we give a brief review of the IFM.
In Section~\ref{section-IFM-with-2DEG}, we explain how to implement the IFM with the 2DEG system.
In Section~\ref{section-shot-noise}, we study the shot noise of the electric current in the interferometer of Kwiat et al.
with an imperfect absorber.
In Section~\ref{section-changing-absorption-coefficient},
changing the absorption coefficient of the object by adjusting the voltage applied to the tip of the tunnelling microscope,
we evaluate the success probability of the IFM.
In Section~\ref{section-discussion},
we give a brief discussion.

\section{\label{section-review-IFM}A brief review of the IFM}
In this section, we explain the scheme of the IFM proposed by Kwiat et al.
This section is a short review of References~\cite{Elitzur1993,Vaidman2001,Kwiat1995}.
The facts described in this section are utilized
in References~\cite{Azuma2003,Azuma2004,Azuma2006}.

\begin{figure}
\begin{center}
\includegraphics{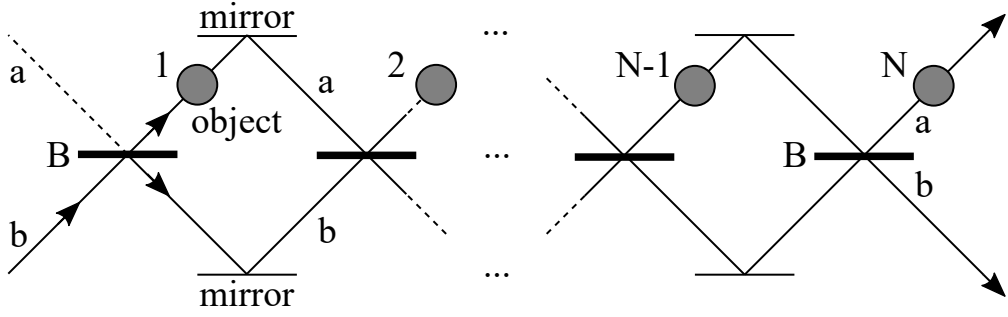}
\end{center}
\caption{The interferometer of Kwiat et al. for the IFM.}
\label{Figure01}
\end{figure}

We start by considering an interferometer that consists of $N$ beam splitters
as shown in Figure~\ref{Figure01}.
The beam splitters divide the interferometer into two parts,
the upper and lower halves.
We write the upper and lower paths as $a$ and $b$ in the interferometer, respectively.
We can regard Figure~\ref{Figure01} as a series of joined Mach-Zehnder interferometers.
We describe a state where one photon travels on the paths $a$ as $|1\rangle_{a}$
and a state where no photon travels on the paths $a$ as $|0\rangle_{a}$.
A similar notation is applied to the paths $b$, as well.
The beam splitter $B$ in Figure~\ref{Figure01}
works as follows:
\begin{equation}
B:
\left\{
\begin{array}{lll}
|1\rangle_{a}|0\rangle_{b} & \to &
\cos\theta|1\rangle_{a}|0\rangle_{b}-\sin\theta|0\rangle_{a}|1\rangle_{b}, \\
|0\rangle_{a}|1\rangle_{b} & \to &
\sin\theta|1\rangle_{a}|0\rangle_{b}+\cos\theta|0\rangle_{a}|1\rangle_{b}.
\end{array}
\right.
\label{definition-beam-splitter-B-01}
\end{equation}
The transmissivity and reflectivity of the beam splitter $B$ are given by
${\cal T}=\sin^{2}\theta$ and ${\cal R}=\cos^{2}\theta$,
respectively.

Let us put a photon into the lower left port of $b$ in Figure~\ref{Figure01}.
If an object does not exist on the paths,
the state of the photon coming from the $k$th beam splitter is given by
\begin{equation}
\sin k\theta|1\rangle_{a}|0\rangle_{b}
+
\cos k\theta|0\rangle_{a}|1\rangle_{b}
\quad\quad
\mbox{for $k=0,1,...,N$.}
\label{photon-state-kth-beam-splitter-01}
\end{equation}
If we set $\theta=\pi/(2N)$,
the photon coming from the $N$th beam splitter flies away to the upper right port of $a$
with probability unity.

Next, we consider the case where a photon-absorbing object exists on the paths $a$.
We assume that the object is put on every path $a$ that comes from each beam splitter
and all of these $N$ objects are the same one.
The photon put into the lower left port of $b$ cannot fly away to the upper right port of $a$
because the object absorbs it.
If the incident photon flies away to the lower right port of $b$ in Figure~\ref{Figure01},
it has not traveled through the paths $a$ in the interferometer.
Thus, a probability that the photon flies away to the lower right port of $b$ is
equal to a product of the reflectivities of the beam splitters,
and it is given by $P=\cos^{2N}\theta$.
In the limit of $N\to\infty$,
$P$ approaches unity asymptotically as follows:
\begin{eqnarray}
\lim_{N\to\infty}P
&=&
\lim_{N\to\infty}\cos^{2N}\Bigl(\frac{\pi}{2N}\Bigr) \nonumber \\
&=&
\lim_{N\to\infty}
[1-\frac{\pi^{2}}{4N}+O\Bigl(\frac{1}{N^{2}}\Bigr)] \nonumber \\
&=&1.
\label{large-N-limit-P-01}
\end{eqnarray}

From the above considerations,
in the large $N$ limit,
we conclude as follows:
(1)
If there is no absorbing object in the interferometer,
the photon flies away to the upper right port of $a$.
(2)
If there is the absorbing object in the interferometer,
the photon flies away to the lower right port of $b$.
Hence, we can examine whether or not the absorbing object exists in the interferometer.

\section{\label{section-IFM-with-2DEG}Implementation of the IFM with the 2DEG}
To implement the IFM with the 2DEG
buried in the
GaAs/AlGaAs heterojunction,
we have to prepare the following three elements:
a single-electron source,
a beam splitter,
and an absorbing object.
In this section, we consider how to construct these three elements in the mesoscopic system.

\begin{figure}
\begin{center}
\includegraphics{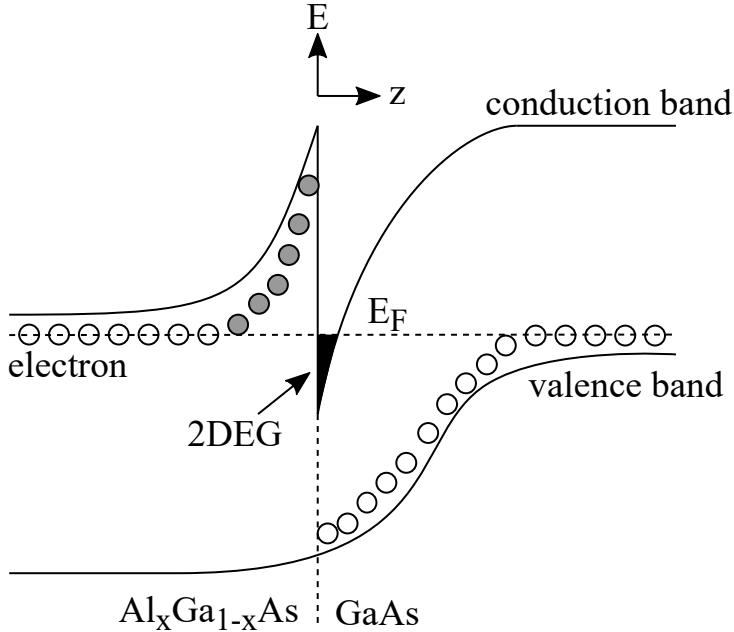}
\end{center}
\caption{A schematic band diagram of the GaAs/AlGaAs heterostructure.}
\label{Figure02}
\end{figure}

Before going into details of the mesoscopic devices,
we study the 2DEG that offers a system of high mobility electrons~\cite{Mimura1980,Ferry1997}.
As shown in Figure~\ref{Figure02},
the heterojunction generated with layers of Si-doped n-type AlGaAs and GaAs
causes a quantum well, into which mobile electrons supplied by n-type AlGaAs drop.
Thus, a thin depleted AlGaAs layer arises.
Because the quantum well forms a steep valley with narrow width
$\Delta z\simeq 8.5\times 10^{-9}$ m
and it is shorter than the Fermi wavelength of the mobile electrons,
they are confined in the $xy$ plane.
Moreover, the valley is located in the GaAs layer where no dopant impurities exist,
so that the confined electrons move without scattering and we can observe the ballistic transport.

First,
we discuss how to realize the single-electron source and the beam splitter.
Implementation of these devices has been achieved already in the field of the electron quantum optics.
In Reference~\cite{Feve2007,Mahe2010},
experiments of an on-demand coherent single-electron emitter are performed.
In these demonstrations, the source is made of a quantum dot coupled to the 2DEG by a quantum point contact (QPC)~\cite{vanWess1988}.
Applying a voltage step to a capacitively coupled gate,
we can increase the dot potential and let an electron occupying the highest energy level of the dot be emitted.
Putting the dot potential in its initial level again,
we can let another electron enter into the dot in compensation with leaving a hole in the Fermi sea.
Repeating this process,
we can let electrons be emitted at constant intervals.
An energy width of a wave packet of the emitted electron is in inverse proportion to the tunnelling time.
This on-demand single electron source works at a high magnetic field,
namely in the quantum Hall regime with no spin degeneracy.

In Reference~\cite{Bocquillon2012},
the electronic beam splitter is made of the QPC and
the Hanbury Brown and Twiss experiment is performed.
In Reference~\cite{Roussel2017},
how to construct quantum signal processor with the electron quantum optics is discussed comprehensively.

Remembering that the electron's averaged relaxation time in GaAs is given by
$\tau\simeq 3.81\times 10^{-11}$ s,
we put the time interval for emissions of single electrons at $\tau_{\mbox{\scriptsize c}}=10^{-9}$ s,
for instance.
Then,
we can obtain the electric current
$I=e/\tau_{\mbox{\scriptsize c}}\simeq 1.60\times 10^{-10}$ A.

Here, we emphasize that generation of single photons at regular intervals is very difficult in general.
An experimental demonstration of emitting single photons under sub-Poisson photon statistics,
such as the photon gun,
is one of the most active topics in the field of quantum information processing~\cite{Kuhn1999,Santori2001,Pelton2002,Keller2004}.
In contrast, turning our eyes to the quantum transport in the mesoscopic systems,
we can construct a single-electron source
with the quantum dot coupled to the 2DEG as mentioned above,
the single-electron tunnelling oscillation given rise to by the Coulomb blockade~\cite{Ferry1997,Averin1986,Ueda1990,Furusaki1992},
and so on.

Employing the QPCs as beam splitters for electrons,
many researchers had built interferometers for the 2DEG system freely
before the concept of the electron quantum optics started to exist.
In References~\cite{Liu1998,Oliver1999,Neder2007a,Neder2007b},
experiments of interference with electrons are performed,
using techniques of the mesoscopic physics.

\begin{figure}
\begin{center}
\includegraphics{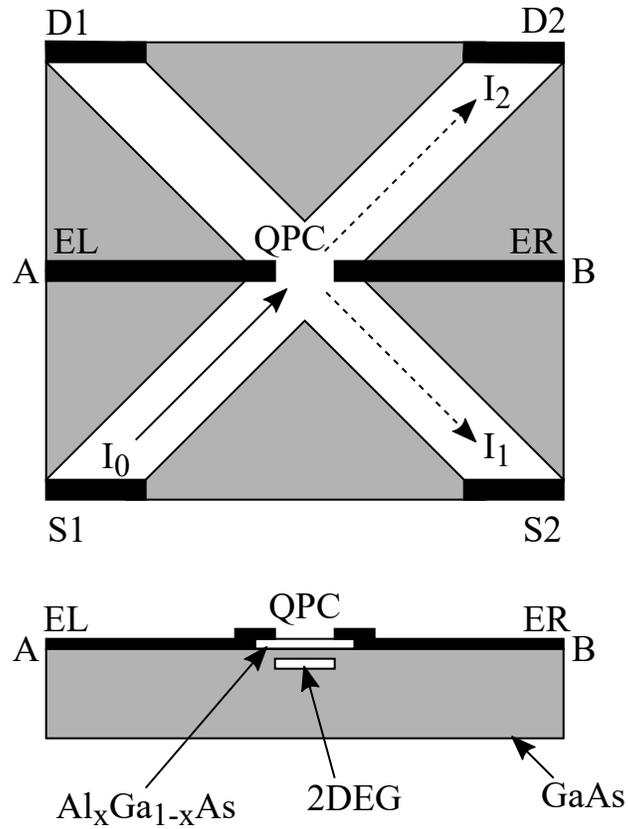}
\end{center}
\caption{A schematic diagram of the beam splitter for the electron made of the QPC
and its cross section between A and B.}
\label{Figure03}
\end{figure}

In Figure~\ref{Figure03},
we show a diagram of the beam splitter made of the QPC.
We apply the drain-to-source voltage $V(>0)$ between drains D1, D2 and sources S1, S2
and the gate bias $V_{\mbox{\scriptsize g}}(<0)$ to the electrodes EL and ER.
The number of channels of the 2DEG varies according to the voltage $V_{g}$.
Burying impurity atoms at the QPC,
we can let the ballistic electrons be scattered with the transmissivity ${\cal T}=\sin^{2}\theta$ and the reflectivity ${\cal R}=\cos^{2}\theta$.
Then,
the QPC works as a beam splitter for electrons.
We examine the shot noise of the electric current flowing through the interferometer of Kwiat et al. that is made of this beam splitter
with the absorbing object
in Section~\ref{section-shot-noise}.

Second, we think about the absorbing object realized by the tunnelling microscope~\cite{Binnig1982,Topinka2001,Cheng2010,Yan2012}.
In Reference~\cite{Topinka2001},
the 2DEG is confined $57$ nm below the surface of the heterojunction.
Letting the distance between a tip and the AlGaAs surface
be stabilized within $1.0\times 10^{-10}$ m around,
we can detect the tunnelling current,
so that we can regard the tip of the tunnelling microscope as the absorbing object.
We define a variable $s$ as a tunnelling distance between the tip and the 2DEG layer.
The tunnelling current feels the potential barriers of the $57$ nm thick AlGaAs and the $0.1$ nm thick vacuum.
Thus, we can approximately consider that the tunnelling current travels only through the AlGaAs layer.

Using the WKB approximation for evaluating the tunnelling current $J(s)$
of the one-dimensional potential barrier,
we obtain
\begin{equation}
J(s)
\simeq
J_{0}e^{-\kappa s},
\label{tunneling-current-formula-01}
\end{equation}
\begin{equation}
\kappa
=
\frac{2}{\hbar}\sqrt{2m^{*}\Bigl(\langle \Phi\rangle-\frac{e|V|}{2}\Bigr)},
\label{tunneling-current-formula-02}
\end{equation}
where $\langle \Phi\rangle$ is a mean of the potential barrier (the work function) of AlGaAs
and $V$ is the negative voltage applied to the tip~\cite{Ferry1997,Schiff1968,Tersoff1985}.
We let the effective mass of the electron be given by $m^{*}=0.067 m_{\mbox{\scriptsize e}}$.
In Equations~(\ref{tunneling-current-formula-01}) and (\ref{tunneling-current-formula-02}), we assume
$\langle \Phi\rangle-e|V|/2>0$,
and $\kappa>0$ holds.
For AlGaAs,
$\langle \Phi\rangle$ is on the order of $5.0$ eV.
Because the 2DEG is buried $57$ nm below the surface as mentioned above,
we let $s\sim 6.0\times 10^{-8}$ m.
If we set $\langle\Phi\rangle-e|V|/2=2.0\times 10^{-4}$ eV,
we obtain $\kappa=3.75\times 10^{7}$ m${}^{-1}$
and $\kappa s=2.25$.
Varying the applied voltage $V$ for the tip of the tunnelling microscope,
we can adjust the probability that the tip captures the electron.

\section{\label{section-shot-noise}The shot noise caused by the QPCs in the interferometer}
In this section,
we examine the shot noise of the electric current induced by the QPCs in the interferometer of Kwiat et al.
Calculating the electric current,
we assume that the interaction between the absorbing object and the interrogating electron is imperfect.
This defect affects the shot noise very much.

\begin{figure}
\begin{center}
\includegraphics{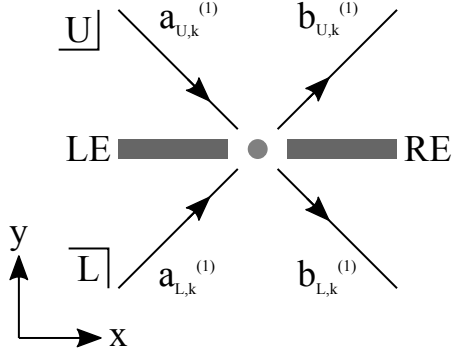}
\end{center}
\caption{A schematic diagram of the first beam splitter that is made of the QPC.
LE and RE represent the left and right electrodes, respectively.
We describe the upper region as U and the lower region as L,
so that the LE and RE form the boundary line between the regions U and L in this figure.
At the QPC,
we bury impurity atoms to cause scattering of the ballistic electrons.
We write annihilation operators of incoming and outgoing electrons around the QPC in the region U as
$a_{\mbox{\scriptsize U},k}^{(1)}$ and $b_{\mbox{\scriptsize U},k}^{(1)}$, respectively.
Similarly,
we write annihilation operators of incoming and outgoing electrons around the QPC in the region L as
$a_{\mbox{\scriptsize L},k}^{(1)}$ and $b_{\mbox{\scriptsize L},k}^{(1)}$, respectively.}
\label{Figure04}
\end{figure}

In Figure~\ref{Figure04},
we show the first beam splitter of the interferometer.
The ballistic electron travels freely in the $x$-direction from left to right with a constant velocity.
By contrast,
it is scattered by impurity atoms located at the QPC in the $y$-direction.
We name the upper and lower regions U and L respectively,
so that the left and right electrodes divide the beam splitter into two halves U and L.
We can regard behaviour of the ballistic electrons in the regions U and L as a one-dimensional quantum system practically.

Only electrons near to the Fermi surface can contribute to transport properties.
Thus,
we only have to consider the ballistic electrons whose wavenumbers are approximately equal to $\pm k_{\mbox{\scriptsize F}}$,
where $k_{\mbox{\scriptsize F}}=\sqrt{2 m^{*}E_{\mbox{\scriptsize F}}}/\hbar$ and $E_{\mbox{\scriptsize F}}$ is the Fermi energy.
In the region of L, we define $a_{k}$ and $a_{k}^{\dagger}$ as annihilation and creation operators of the electron with $k\simeq k_{\mbox{\scriptsize F}}$,
which moves towards the positive direction of the $y$-axis.
Similarly,
we define $b_{k}$ and $b_{k}^{\dagger}$ as annihilation and creation operators of the electron with $k\simeq -k_{\mbox{\scriptsize F}}$,
which moves towards the negative direction of the $y$-axis.
In the region of U, we reverse the $y$-axis.
Thus, for the first beam splitter,
we write annihilation operators of incoming electrons to the QPC in the regions L and U as
$a_{\mbox{\scriptsize L},k}^{(1)}$ and $a_{\mbox{\scriptsize U},k}^{(1)}$, respectively.
Similarly,
we write annihilation operators of outgoing electrons from the QPC in the regions L and U as
$b_{\mbox{\scriptsize L},k}^{(1)}$ and $b_{\mbox{\scriptsize U},k}^{(1)}$, respectively.

Because there are the impurity atoms at the QPC,
the ballistic electrons are scattered and the scattering matrix is given by
\begin{equation}
\left(
\begin{array}{c}
b_{\mbox{\scriptsize U},k}^{(1)} \\
b_{\mbox{\scriptsize L},k}^{(1)} \\
\end{array}
\right)
=
B
\left(
\begin{array}{c}
a_{\mbox{\scriptsize U},k}^{(1)} \\
a_{\mbox{\scriptsize L},k}^{(1)} \\
\end{array}
\right),
\end{equation}
\begin{equation}
B
=
\left(
\begin{array}{cc}
r' & t \\
t' & r \\
\end{array}
\right),
\label{scattering-matrix-01}
\end{equation}
$r=r'=\cos\theta$, and $t=-t'=\sin\theta$.
This implies that the transmissivity and reflectivity of the beam splitter shown in Figure~\ref{Figure04}
are given by
${\cal T}=\sin^{2}\theta$ and ${\cal R}=\cos^{2}\theta$, respectively.
We note that the matrix $B$ is consistent with Equation~(\ref{definition-beam-splitter-B-01}).

Here, we assume that the interaction between the absorbing object and the interrogating electron is imperfect,
so that the electron is absorbed with probability $(1-\eta)$ and it passes by the object without being absorbed with probability $\eta$
when it makes a close approach to the object.
Defining the matrix,
\begin{equation}
A
=
\left(
\begin{array}{cc}
\sqrt{\eta} & 0 \\
0 & 1 \\
\end{array}
\right),
\label{absorption-matrix-01}
\end{equation}
we can write down annihilation operators of outgoing electrons from $N$th beam splitter as
\begin{equation}
\left(
\begin{array}{c}
b_{\mbox{\scriptsize U},k}^{(N)} \\
b_{\mbox{\scriptsize L},k}^{(N)} \\
\end{array}
\right)
=
S
\left(
\begin{array}{c}
a_{\mbox{\scriptsize U},k}^{(1)} \\
a_{\mbox{\scriptsize L},k}^{(1)} \\
\end{array}
\right),
\end{equation}
where $S=(BA)^{N}$.
An explicit form of $S$ is given in Reference~\cite{Azuma2006}.

An operator of the electric current detected on the lower right port in the interferometer of Kwiat et al. shown in Figure~\ref{Figure01} is given by
\begin{eqnarray}
I_{\mbox{\scriptsize L}b}(t)
&=&
\frac{ev_{\mbox{\scriptsize F}}}{L}
\sum_{k,k'}
b_{\mbox{\scriptsize L},k}^{(N)\dagger}
b_{\mbox{\scriptsize L},k'}^{(N)}
\exp[i(\varepsilon_{k}-\varepsilon_{k'})t] \nonumber \\
&=&
\frac{ev_{\mbox{\scriptsize F}}}{L}
\sum_{k,k'}
\sum_{\alpha,\beta\in\{\mbox{\scriptsize U},\mbox{\scriptsize L}\}}
(S_{\mbox{\scriptsize L}\alpha}a_{\alpha,k}^{(1)})^{\dagger}
S_{\mbox{\scriptsize L}\beta}a_{\beta,k'}^{(1)}
\exp[i(\varepsilon_{k}-\varepsilon_{k'})t].
\label{electric-current-operator-01}
\end{eqnarray}
In Equation~(\ref{electric-current-operator-01}),
we introduce the length $L$ for a periodic boundary condition of a wave function of the electron in a box.

Here,
we assume that the incident electrons are in thermal equilibrium,
\begin{equation}
\langle
a_{\alpha,k}^{(1)\dagger}a_{\beta,k'}^{(1)}
\rangle
=
\delta_{\alpha\beta}
\delta_{kk'}
f_{\alpha}(k),
\label{thermal-equilibrium-01}
\end{equation}
where $f_{\alpha}(k)$ is the Fermi-Dirac distribution function.
Strictly speaking,
the incident electrons are generated by the on-demand single electron emitter
and they form a discrete electric current far from thermal equilibrium.
However,
for the sake of simplicity,
we assume Equation~(\ref{thermal-equilibrium-01}).

Now, we calculate a noise spectral density $S(\omega)$ at the frequency $\omega=0$,
\begin{eqnarray}
S(0)
&=&
2\int_{-\infty}^{\infty}dt
[
\langle
I_{\mbox{\scriptsize L}b}(t)
I_{\mbox{\scriptsize L}b}(0)
\rangle
-
\langle
I_{\mbox{\scriptsize L}b}(t)
\rangle
\langle
I_{\mbox{\scriptsize L}b}(0)
\rangle
] \nonumber \\
&=&
2\int_{-\infty}^{\infty}dt
\Bigl(
\frac{ev_{\mbox{\scriptsize F}}}{L}
\Bigr)^{2}
\sum_{k,k',k'',k'''}
\sum_{\alpha,\beta,\alpha'\beta'\in\{\mbox{\scriptsize U},\mbox{\scriptsize L}\}}
S_{\mbox{\scriptsize L}\alpha}^{*}
S_{\mbox{\scriptsize L}\beta}
S_{\mbox{\scriptsize L}\alpha'}^{*}
S_{\mbox{\scriptsize L}\beta'} \nonumber \\
&&\quad\quad
\times
[
\langle
a_{\alpha k}^{(1)\dagger}
a_{\beta k'}^{(1)}
a_{\alpha' k''}^{(1)\dagger}
a_{\beta' k'''}^{(1)}
\rangle
-
\langle
a_{\alpha k}^{(1)\dagger}
a_{\beta k'}^{(1)}
\rangle
\langle
a_{\alpha' k''}^{(1)\dagger}
a_{\beta' k'''}^{(1)}
\rangle
]
\exp[i(\varepsilon_{k}-\varepsilon_{k'})t].
\end{eqnarray}
Here,
we use the Bloch-de Dominicis theorem,
\begin{eqnarray}
\langle
a_{\alpha k}^{(1)\dagger}
a_{\beta k'}^{(1)}
a_{\alpha' k''}^{(1)\dagger}
a_{\beta' k'''}^{(1)}
\rangle
&=&
\langle
a_{\alpha k}^{(1)\dagger}
a_{\beta k'}^{(1)}
\rangle
\langle
a_{\alpha' k''}^{(1)\dagger}
a_{\beta' k'''}^{(1)}
\rangle
-
\langle
a_{\alpha k}^{(1)\dagger}
a_{\alpha' k''}^{(1)\dagger}
\rangle
\langle
a_{\beta k'}^{(1)}
a_{\beta' k'''}^{(1)}
\rangle \nonumber \\
&&
\quad\quad
+
\langle
a_{\alpha k}^{(1)\dagger}
a_{\beta' k'''}^{(1)}
\rangle
\langle
a_{\beta k'}^{(1)}
a_{\alpha' k''}^{(1)\dagger}
\rangle,
\end{eqnarray}
and pay attention to
$\langle
a_{\alpha k}^{(1)\dagger}
a_{\alpha' k''}^{(1)\dagger}
\rangle
\langle
a_{\beta k'}^{(1)}
a_{\beta' k'''}^{(1)}
\rangle
=0$.

Then,
we obtain
\begin{equation}
S(0)
=
2\int_{-\infty}^{\infty}dt
\Bigl(
\frac{ev_{\mbox{\scriptsize F}}}{L}
\Bigr)^{2}
\sum_{k,k'}
\sum_{\alpha,\beta\in\{\mbox{\scriptsize U},\mbox{\scriptsize L}\}}
S_{\mbox{\scriptsize L}\alpha}^{*}
S_{\mbox{\scriptsize L}\beta}
S_{\mbox{\scriptsize L}\beta}^{*}
S_{\mbox{\scriptsize L}\alpha}
f_{\alpha}(k)[1-f_{\beta}(k')]
\exp[i(\varepsilon_{k}-\varepsilon_{k'})t].
\end{equation}
Next,
we replace $(1/L)\sum_{k}$ with $[1/(2\pi)]\int dk$.
Moreover,
because the kinetic energy of the incident electron is given by
$\varepsilon_{k}=v_{\mbox{\scriptsize F}}(k-k_{\mbox{\scriptsize F}})$,
we replace $\int dk$ with $(1/v_{\mbox{\scriptsize F}})\int d\varepsilon$.
Thus,
we obtain
\begin{equation}
S(0)
=
\frac{e^{2}}{\pi}
\int_{-\infty}^{\infty}
d\varepsilon
\sum_{\alpha,\beta\in\{\mbox{\scriptsize U},\mbox{\scriptsize L}\}}
S_{\mbox{\scriptsize L}\alpha}^{*}
S_{\mbox{\scriptsize L}\beta}
S_{\mbox{\scriptsize L}\beta}^{*}
S_{\mbox{\scriptsize L}\alpha}
f_{\alpha}(\varepsilon)[1-f_{\beta}(\varepsilon)],
\end{equation}
where we make use of
\begin{equation}
\int_{-\infty}^{\infty}dt
\exp[i(\varepsilon-\varepsilon')t]=2\pi\delta(\varepsilon-\varepsilon').
\end{equation}

We assume that an electric potential $V(<0)$ is applied to the region L and the temperature of the system is given by
$T=0$ K.
Then,
the Fermi-Dirac distribution functions in the regions L and U are given by
\begin{eqnarray}
f_{\mbox{\scriptsize L}}(\varepsilon)&=&\Theta(-\varepsilon+e|V|), \nonumber \\
f_{\mbox{\scriptsize U}}(\varepsilon)&=&\Theta(-\varepsilon),
\end{eqnarray}
where $\Theta(x)$ is the Heaviside step function.
If and only if $\alpha=L$ and $\beta=U$,
$f_{\alpha}(\varepsilon)[1-f_{\beta}(\varepsilon)]$ can take a non-zero value,
such that
\begin{equation}
f_{\mbox{\scriptsize L}}(\varepsilon)[1-f_{\mbox{\scriptsize U}}(\varepsilon)]
=
\left\{
\begin{array}{ll}
1 & 0<\varepsilon<e|V|, \\
0 & \mbox{otherwise.} \\
\end{array}
\right.
\end{equation}
Finally,
we obtain
\begin{equation}
S(0)
=
\frac{e^{2}}{\pi}
|S_{\mbox{\scriptsize L}\mbox{\scriptsize L}}|^{2}
|S_{\mbox{\scriptsize L}\mbox{\scriptsize U}}|^{2}
e|V|.
\end{equation}

\begin{figure}
\begin{center}
\includegraphics{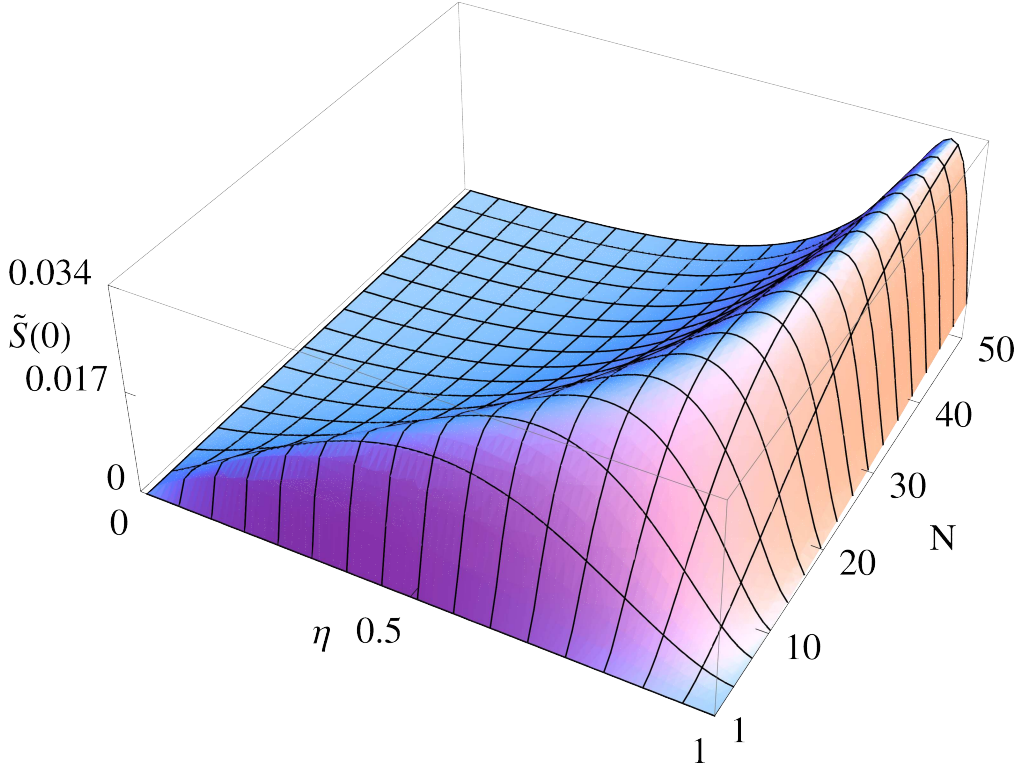}
\end{center}
\caption{A graph of $\tilde{S}(0)=\pi S(0)/(e^{3}|V|)$,
where $S(0)$ is the noise spectral density at the frequency $\omega=0$,
as a function of $(N,\eta)$
for $1\leq N\leq 50$ and $0\leq \eta \leq 1$.
$N$, $\eta$, and $\tilde{S}(0)$ are dimensionless.}
\label{Figure05}
\end{figure}

We show a graph of $\tilde{S}(0)=\pi S(0)/(e^{3}|V|)$ as a function of $(N,\eta)$ in Figure~\ref{Figure05},
where $\theta=\pi/(2N)$, $1\leq N\leq 50$, and $0\leq \eta \leq 1$.
When $\eta=0$,
quantum interference does not occur in the interferometer,
so that the shot noise is never induced and we obtain $S(0)=\tilde{S}(0)=0$.

\section{\label{section-changing-absorption-coefficient}Changing the absorption coefficient
of the object by adjusting the tip's bias}
In this section, we consider the success probability of the IFM with the imperfect absorber
that is realized with the tunnelling microscope.
The dependence of the success probability of the IFM on a rate at which the object captures the interrogating particle is investigated
in Reference~\cite{Azuma2006}.

Looking at Equations~(\ref{tunneling-current-formula-01}) and (\ref{tunneling-current-formula-02}),
we notice that the tunnelling current is sensitive to a change of the bias of the tip of the tunnelling microscope.
In Section~~\ref{section-shot-noise},
we define $(1-\eta)$ as the absorption coefficient of the object.
From Equations~(\ref{tunneling-current-formula-01}) and (\ref{tunneling-current-formula-02}),
we can write down the variable $\eta$ as a function of
$\Delta W=\langle\Phi\rangle-e|V|/2$,
namely
\begin{equation}
\eta(\Delta W)
=
1
-
\exp[-\kappa(\Delta W)s],
\label{definition-absorption-coefficient-01}
\end{equation}
\begin{equation}
\kappa(\Delta W)
=
(2/\hbar)
\sqrt{2m^{\ast}\Delta W}.
\label{definition-absorption-coefficient-02}
\end{equation}
A probability that an incident electron from the lower left port of $b$ passes through the $N$ beam splitters and is detected in the lower port of $b$
in Figure~\ref{Figure01} is given by
\begin{equation}
P(N,\Delta W)=|\langle \bar{1}|(BA)^{N-1}B|\bar{1}\rangle|^{2},
\label{success-probability-IFM-01}
\end{equation}
where
\begin{equation}
|\bar{0}\rangle
=
|1\rangle_{a}|0\rangle_{b}
=
\left(
\begin{array}{c}
1 \\
0
\end{array}
\right),
\quad\quad
|\bar{1}\rangle
=
|0\rangle_{a}|1\rangle_{b}
=
\left(
\begin{array}{cc}
0 \\
1
\end{array}
\right).
\end{equation}
The matrices $A$ and $B$ are given in Equations~(\ref{scattering-matrix-01}) and (\ref{absorption-matrix-01})
and $\theta=\pi/(2N)$.
Substitution of Equations~(\ref{definition-absorption-coefficient-01}) and (\ref{definition-absorption-coefficient-02})
into Equation~(\ref{success-probability-IFM-01}) yields the function of the probability
$P(N,\Delta W)$.

\begin{figure}
\begin{center}
\includegraphics{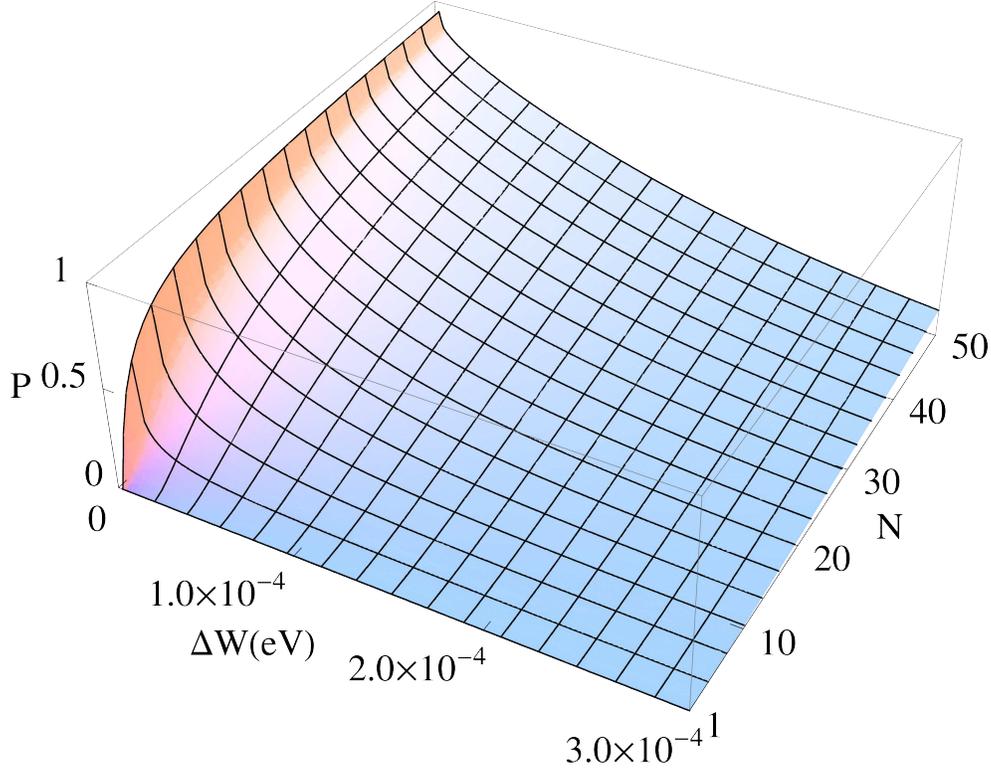}
\end{center}
\caption{A graph of the probability $P(N,\Delta W)$ for $1\leq N\leq 50$ and $0\leq \Delta W\leq 3.0\times 10^{-4}$ eV.
The unit of $\Delta W$ for the graph is the electron volt.
$N$ and $P$ are dimensionless.}
\label{Figure06}
\end{figure}

Here,
we calculate $P(N,\Delta W)$ numerically.
We fix $s=6.0\times 10^{-8}$ m.
We plot $P=P(N,\Delta W)$ for $1\leq N\leq 50$ and $0\leq \Delta W\leq 3.0\times 10^{-4}$ eV in Figure~\ref{Figure06}.
From Figure~\ref{Figure06},
we notice that $P(N,\Delta W)$ is sensitive to a change of $\Delta W$.

\section{\label{section-discussion}Results and discussion}
In this paper, we consider how to implement the IFM with the 2DEG system.
Particularly, we argue that we can realize the absorbing object with the tunnelling microscope
by adjusting the voltage applied between the heterojunction and the tip.
We evaluate the shot noise of the electric current induced by the interferometer of Kwiat et al.
We confirm that the absorption rate of the object affects the shot noise very much.

The probability $P(N,\Delta W)$ given in Equation~(\ref{success-probability-IFM-01})
is proportional to the current observed at the lower right port of $b$ in Figure~\ref{Figure01}.
This implies that the single-electron source realized by the quantum dot coupled to the 2DEG has to generate a constant current.
Thus, the intervals between single-electron emissions are critical and their fluctuation caused by the shot noise is a serious problem.

In general,
generating a source of single electrons is as difficult as constructing a triggered source for single photons.
If we try to build a single-photon gun from laser light naively,
we encounter the following two problems.
The first one is fluctuation of the number of photons emitted together.
Considering a coherent state $|\alpha\rangle$,
we obtain
$\langle\hat{n}\rangle=|\alpha|^{2}$ and $\Delta n^{2}=|\alpha|^{2}$
for
the expectation value and the variance of the photon number operator, respectively.
Thus, as long as we utilize the coherent light as the source of single photons,
we have to worry about photon bunching.
The second one is fluctuation of the time interval for photon emission
because of its Poisson distribution.
Thinking about a single-electron gun,
we can suppress the fluctuation of the number of electrons with the Fermi-Dirac statistics.
However,
a problem with the fluctuation of the time interval of the electron emission remains.
To generate the constant current of the electrons,
we have to apply strictly periodic voltage steps to the quantum dot coupled to the 2DEG as mentioned in Section~\ref{section-IFM-with-2DEG}.

In this paper,
we utilize the tunnelling microscope for absorbing the electrons.
Alternatively,
we can make use of the single electron transistor for capturing interrogating electrons \cite{Ferry1997}.
However,
we prefer the tunnelling microscope to the single electron transistor because of the simplicity of device structures.

As mentioned in Section~\ref{section-introduction},
so far,
many researchers have adopted interrogating photons to perform experiments of the IFM.
However,
the author thinks that the mesoscopic system on the 2DEG offers us
a robust platform for quantum information processing.

\section*{Acknowledgements}
This paper is an extended version of a poster presentation in nc-AFM 2011~\cite{Azuma2011}.

\end{document}